\documentstyle[12pt,amssymb]{article}
\newcommand{\text}[1]{\mathrm{#1}}

\begin{document}

\sloppy

\pagestyle{empty}

\begin{flushright}
CERN-TH.7043/93 \\
DTP/93/84
\end{flushright}

\vspace{\fill}

\begin{center}
{\LARGE\bf Does the W Mass Reconstruction}\\[4mm]
{\LARGE\bf Survive QCD Effects?}\\[25mm]
{\Large Torbj\"orn Sj\"ostrand} \\[3mm]
{\large Theory Division, CERN} \\[1mm]
{\large CH-1211 Geneva 23, Switzerland}\\[5mm]
{\large and} \\[5mm]
{\Large Valery A. Khoze$~*$} \\[3mm]
{\large Department of Physics, University of Durham} \\[1mm]
{\large Durham DH1 3LE, England} \\
\end{center}

\vspace{\fill}

\begin{center}
\bf{Abstract}
\end{center}
\vspace{-0.5\baselineskip}
\noindent
In the hadronic decay mode of a pair of W bosons,
$\text{e}~+\text{e}~- \to \text{W}~+ \text{W}~- \to
\text{q}_1 \overline{\text{q}}_2 \text{q}_3 \overline{\text{q}}_4$,
QCD interference effects can mix up the two colour singlets
$\text{q}_1 \overline{\text{q}}_2$ and
$\text{q}_3 \overline{\text{q}}_4$, i.e. produce hadrons that
cannot be uniquely assigned to either of W$~+$ and W$~-$.
We show that interference is negligible for energetic perturbative
gluon emission, and develop models to help us to estimate the
non-perturbative effects. The total contribution to the systematic
error on the W mass reconstruction may be as large as 40 MeV.

\vspace{\fill}

\noindent
CERN-TH.7043/93 \\
October 1993

\clearpage

\pagestyle{plain}
\setcounter{page}{1}

This letter is concerned with QCD interference effects
that may occur when two unstable particles decay and hadronize
close to each other in space and time. W pair production
is here of particular interest because of its practical
importance and relative simplicity, but many other processes
are also affected. (For further details and references
see Ref. \cite{bigpaper}.)

QCD interference effects between the W$~+$ and
W$~-$ decays undermine the traditional meaning of a
W mass in the process
$\text{e}~+\text{e}~- \to \text{W}~+ \text{W}~- \to
\text{q}_1 \overline{\text{q}}_2 \text{q}_3 \overline{\text{q}}_4$.
Specifically, it is not even in principle possible to
subdivide the hadronic final state into two groups of
particles, one of which is produced by the
$\text{q}_1\overline{\text{q}}_2$
system of the W$~+$ decay and the other by the
$\text{q}_3\overline{\text{q}}_4$ system of the
W$~-$ decay: some particles
originate from the joint action of the two systems.
Since a determination of the W mass is one of the main
objectives of LEP
large the ambiguities can be. A statistical error of
55
the precision of the theoretical predictions should
ideally match or exceed this experimental accuracy.

A complete description of QCD interference effects is
not possible since non-perturbative QCD is not well
understood. The concept of colour reconnection/rearrangement
is therefore useful to quantify effects
(at least in a first approximation).
In a reconnection two original colour singlets
(such as $\text{q}_1\overline{\text{q}}_2$ and
$\text{q}_3\overline{\text{q}}_4$)
are transmuted into two new ones
(such as $\text{q}_1\overline{\text{q}}_4$ and
$\text{q}_3\overline{\text{q}}_2$).
Subsequently each singlet system is assumed to hadronize
independently according to the standard algorithms, which have
been so successful in describing e.g. Z$~0$ decays. Depending on
whether a reconnection has occurred or not, the hadronic final
state is then going to be somewhat different.

The colour reconnection effects were first studied by
Gustafson, Pettersson and Zerwas \cite{GPZ}, but their results
were mainly qualitative and were not targeted on what might
actually be expected at LEP
example of the so-called instantaneous reconnection scenario,
where the alternative colour singlets are immediately formed
and allowed to radiate perturbative gluons.

In order to understand which QCD interference effects can occur
in hadronic W$~+$W$~-$ decays, it is useful to examine the
space--time picture of the process. Consider a typical c.m.
energy of 170 GeV, a W mass $m_{\text{W}} = 80$ GeV and width
$\Gamma_{\text{W}} = 2.08$ GeV.
The averaged (over the W mass distribution)
proper lifetime for a W is $\langle \tau \rangle
\approx (2/3) \hbar/\Gamma_{\text{W}} \approx 0.06$
This gives a mean separation of the two decay vertices
of 0.04
A gluon with an energy $\omega \gg \Gamma_{\text{W}}$
therefore has a wavelength much smaller than the separation
between the W$~+$ and W$~-$ decay vertices, and is
emitted almost incoherently either by the
$\text{q}_1\overline{\text{q}}_2$ system or by
the $\text{q}_3\overline{\text{q}}_4$ one \cite{K4}.
Only fairly soft gluons, $\omega \lesssim \Gamma_{\text{W}}$,
feel the joint action of all four
quark colour charges. On the other hand, the typical
distance scale of hadronization is about 1
than the decay vertex separation. Therefore the hadronization
phase may contain significant interference effects.

In the following, we will first discuss perturbative effects and
subsequently non-perturbative ones. (For a discussion of a possible
interplay between the two stages see Ref. \cite{bigpaper}.)

Until today, perturbative QCD has mainly been applied to systems
of primary partons produced almost simultaneously. The
radiation accompanying such a system can be represented as a
superposition of gauge-invariant terms, in which each external
quark line is uniquely connected to an external antiquark line
of the same colour. The system is thus decomposed into a set of
colourless $\text{q}\overline{\text{q}}$ antennae/dipoles \cite{K2}.
One of the simplest
examples is the celebrated $\text{q}\overline{\text{q}}\text{g}$
system, which (to leading order in
$1/N_C~2$, where $N_C =3$ is the number of colours) is well
approximated by the incoherent sum of two separate antennae,
$\widehat{\text{q}\text{g}}$ and
$\widehat{\text{g}\overline{\text{q}}}$. These dipoles radiate
gluons, which within the perturbative scenario are the principal
sources of multiple hadroproduction.

Neglecting interferences, the
$\text{e}~+\text{e}~- \to \text{W}~+ \text{W}~- \to
\text{q}_1 \overline{\text{q}}_2 \text{q}_3 \overline{\text{q}}_4$
final state can be subdivided into two separate dipoles,
$\widehat{\text{q}_1 \overline{\text{q}}_2}$ and
$\widehat{\text{q}_3 \overline{\text{q}}_4}$.
Each dipole may radiate gluons from a maximum scale $m_{\text{W}}$
downwards. Within the perturbative approach,
colour transmutations can result only from the interferences
between gluons (virtual as well as real) radiated in the W$~+$ and
W$~-$ decays. A colour reconnection then corresponds to radiation,
e.g. from the dipoles $\widehat{\text{q}_1 \overline{\text{q}}_4}$
and $\widehat{\text{q}_3 \overline{\text{q}}_2}$. The emission of a
single primary gluon cannot give interference effects,
by colour conservation, so interference terms only enter in second
order in $\alpha_s$.

The general structure of the results is well illustrated by
the interference between the graph where a gluon
with momentum $k_1$ ($k_2$) is emitted off the
$\widehat{\text{q}_1 \overline{\text{q}}_2}$
($\widehat{\text{q}_3 \overline{\text{q}}_4}$)
dipole and the same graph with $k_1$ and $k_2$ interchanged:
\begin{equation}
\frac{1}{\sigma_{0}} \, \text{d}\sigma~{\text{int}} \simeq
\frac{\text{d}~3 \mbox{\bf k}_1}{\omega_1} \,
\frac{\text{d}~3 \mbox{\bf k}_2}{\omega_2} \,
\left( \frac{C_F \, \alpha_s}{4\pi~2} \right)~2 \,
\frac{1}{N_C~2 -1} \; \chi_{12} \; H(k_1) \, H(k_2) 
\end{equation}
where $C_F = (N_C~2-1)/(2 N_C) = 4/3$.
We proceed to comment on the non-trivial factors in this expression.

The interference is suppressed by $1/(N_C~2-1) = 1/8$ as compared
to the total rate of double primary gluon emissions. This is a
result of the ratio of the corresponding colour traces.

The so-called profile function $\chi_{12}$ \cite{K4,K5} controls
decay--decay interferences. It quantifies the overlap of the W
propagators in the interfering Feynman diagrams. Near the
W$~+$W$~-$ pair threshold, $\chi_{12}$ simplifies to
\begin{equation}
\chi_{12} \approx \frac{\Gamma_{\text{W}}~2}{\Gamma_{\text{W}}~2 +
(\omega_1 - \omega_2)~2} 
\end{equation}
Other interferences (real or virtual) are described by somewhat
different expressions,
e.g. with $\omega_1 - \omega_2 \to \omega_1 + \omega_2$, but
have the same general properties.
The profile functions cut down the phase space available for
gluon emissions with $\omega \gtrsim \Gamma_{\text{W}}$
by the alternative quark pairs. (We can neglect the contribution
from kinematical configurations with
\mbox{$\omega_1, \omega_2 \gg \Gamma_{\text{W}}$},
\mbox{$|\omega_1 - \omega_2| \lesssim \Gamma_{\text{W}}$}
since the corresponding phase-space volume is small.)
The possibility for the reconnected systems to develop QCD cascades
is thus reduced, i.e. the dipoles are almost sterile.

The radiation pattern $H(k)$ is given by
\begin{equation}
H(k) = \widehat{\text{q}_1 \overline{\text{q}}_4} +
\widehat{\text{q}_3 \overline{\text{q}}_2} -
\widehat{\text{q}_1 \text{q}_3} -
\widehat{\overline{\text{q}}_2 \overline{\text{q}}_4} 
\end{equation}
where the radiation antennae are \cite{K2}
\begin{equation}
\widehat{ij} = \frac{(p_i \cdot p_j)}{(p_i \cdot k)(p_j \cdot k)}
\end{equation}
In addition to the two dipoles
$\widehat{\text{q}_1 \overline{\text{q}}_4}$
and $\widehat{\text{q}_3 \overline{\text{q}}_2}$,
which may be interpreted in
terms of reconnected colour singlets, one finds two
other terms, $\widehat{\text{q}_1 \text{q}_3}$ and
$\widehat{\overline{\text{q}}_2 \overline{\text{q}}_4}$,
which come in with a negative sign. The signs represent the
attractive and repulsive forces between quarks and antiquarks
\cite{K2,K8}.

It should be emphasized that, analogously to other
colour-suppressed interference phenomena,
rearrangement can be viewed only on a completely
inclusive basis, when all the antennae are simultaneously
active in the particle production. The very fact that the
reconnection pieces are not positive-definite reflects their
wave interference nature. Therefore the effects of reconnected
almost sterile cascades should
appear on top of a dominant background generated by the
ordinary-looking no-reconnection dipoles
$\widehat{\text{q}_1 \overline{\text{q}}_2}$
and $\widehat{\text{q}_3 \overline{\text{q}}_4}$.

Summing up the above discussion, it can be concluded that
perturbative colour reconnection phenomena are suppressed,
firstly because of the overall factor $\alpha_s~2/(N_C~2-1)$,
and secondly because the rearranged dipoles can only radiate
gluons with energies $\omega \lesssim \Gamma_{\text{W}}$.
Only a few low-energy particles should therefore be affected,
$\Delta N~{\text{recon}} / N~{\text{no-recon}} \lesssim
{\cal O}(10~{-2})$.

We now turn to the possibility of reconnection
occurring as a part of the non-perturbative hadronization
phase. Since hadronization is not understood from first principles,
this requires model building rather than exact calculations. We will
use the standard Lund string fragmentation model \cite{Lund}
as a starting point, but have to extend it considerably.
The string is here to be viewed as a Lorentz covariant
representation of a linear confinement field.

The string description is entirely
probabilistic, i.e. any negative-sign interference effects
are absent. This means that the original colour singlets
$\text{q}_1\overline{\text{q}}_2$ and
$\text{q}_3\overline{\text{q}}_4$
may transmute to new singlets
$\text{q}_1\overline{\text{q}}_4$ and
$\text{q}_3\overline{\text{q}}_2$,
but that any effects e.g.\ of
$\widehat{\text{q}_1\text{q}_3}$ or
$\widehat{\overline{\text{q}}_2\overline{\text{q}}_4}$
dipoles are absent. In this
respect, the non-perturbative discussion is more limited in
outlook than the perturbative one above. However, note that
dipoles such as $\widehat{\text{q}_1\text{q}_3}$ do not correspond
to colour singlets, and can therefore not survive in the
long-distance limit of the theory, i.e. they have to disappear
in the hadronization phase.

The imagined time sequence is the following. The W$~+$ and
W$~-$ fly apart from their common production vertex and decay
at some distance. Around each of these decay vertices, a
perturbative parton shower evolves from an original
$\text{q}\overline{\text{q}}$
pair. The typical distance that a virtual parton (of mass
$m \sim 10$
hadronic final state) travels before branching is comparable with
the average W$~+$W$~-$ separation, but shorter than the
fragmentation time. Each W can therefore effectively be viewed
as instantaneously decaying into a string spanned between the
partons, from a quark end via a number of
intermediate gluons to the antiquark end. The strings expand,
both transversely and longitudinally, at a speed limited by that
of light. They eventually fragment into hadrons and disappear.
Before that time, however, the string from the W$~+$ and the one
from the W$~-$ may overlap. If so, there is some probability for
a colour reconnection to occur in the overlap region. The
fragmentation process is then modified.

The Lund string model does not constrain the nature of the string
fully. At one extreme, the string may be viewed as an elongated bag,
i.e.\ as a flux tube without any pronounced internal structure.
At the other extreme, the string contains a very thin core, a vortex
line, which carries all the topological information, while the energy is
distributed over a larger surrounding region. The latter alternative
is the chromoelectric analogue to the magnetic flux lines in a type II
superconductor, whereas the former one is more akin to the structure
of a type I superconductor. We use them as starting points for
two contrasting approaches, with nomenclature inspired by the
superconductor analogy.

In scenario
space--time volume over which the W$~+$ and W$~-$
strings overlap, with saturation at unit probability.
This probability is calculated as
follows. In the rest frame of a string piece expanding along the
$\pm z$ direction, the colour field strength is assumed
to be given by
\begin{equation}
\Omega(\mbox{\bf x},t) =
\exp \left\{ - (x~2 + y~2)/2r_{\text{had}}~2 \right\}
\; \theta(t - |\mbox{\bf x}|) \;
\exp \left\{ - (t~2 - z~2)/\tau_{\text{frag}}~2 \right\} 
\end{equation}
The first factor gives a Gaussian fall-off in the transverse
directions, with a string width $r_{\text{had}} \approx 0.5$
of typical hadronic dimensions. The time retardation factor
$\theta(t - |\mbox{\bf x}|)$ ensures that information on the decay of
the W spreads outwards with the speed of light. The last factor
gives the probability that the string has not yet fragmented at
a given proper time along the string axis, with
$\tau_{\text{frag}} \approx 1.5$
from the W$~+$ decay, this field strength has to be appropriately
rotated, boosted and displaced to the W$~+$ decay vertex.
In addition, since the W$~+$ string can be made up of many pieces,
the string field strength $\Omega_{\text{max}}~+(\mbox{\bf x},t)$
is defined as the maximum of all the contributing $\Omega~+$'s
in the given point. The probability for a reconnection to occur
is now given by
\begin{equation}
{\cal P}_{\text{recon}} =  1 - \exp \left( - k_{\text{I}}
\int \text{d}~3\mbox{\bf x} \, \text{d} t \;
\Omega_{\text{max}}~+(\mbox{\bf x},t) \,
\Omega_{\text{max}}~-(\mbox{\bf x},t) \right) 
\end{equation}
where $k_{\text{I}}$ is a free parameter. If a reconnection
occurs, the space--time point for this reconnection is selected
according to the differential probability
$\Omega_{\text{max}}~+(\mbox{\bf x},t) \,
\Omega_{\text{max}}~-(\mbox{\bf x},t)$.
This defines the string pieces involved
and the new colour singlets.

In scenario II it is assumed that reconnections can only
take place when the core regions of two string pieces cross
each other. This means that the transverse extent of
strings can be neglected, which leads to considerable
simplifications compared with the previous scenario.
The position of a string piece at time $t$ is described
by a one-parameter set $\mbox{\bf x}(t,\alpha)$, where
$0 \leq \alpha \leq 1$ is used to denote the position along the
string. To find whether two string pieces $i$ and $j$ from the
W$~+$ and W$~-$ decays cross, it is sufficient to solve the
equation system $\mbox{\bf x}_i~+(t , \alpha~+) =
\mbox{\bf x}_j~-(t , \alpha~-)$ and to check that this
(unique) solution is in the physically allowed
domain. Further, it is required
that neither string piece has had time to fragment, which gives
two extra suppression factors of the form
$\exp \{ - \tau~2/\tau_{\text{frag}}~2 \}$,
with $\tau$ the proper lifetime of each string piece at the
point of crossing, i.e. as in scenario
string crossings, only the one that occurs first is retained.

Both scenarios are implemented in a detailed simulation of the
full process of W$~{\pm}$ production and decay, parton shower
evolution and hadronization \cite{Pythia}. It is therefore
possible to assess any experimental consequences for an ideal
detector.

The reconnection probability is predicted in
scenario
the possibility to vary the baseline model in a few respects.
Scenario
$k_{\text{I}}$. We have chosen $k_{\text{I}}$ to give
an average ${\cal P}_{\text{recon}} \approx 0.35$ at 170 GeV,
as is predicted in scenario

The resulting c.m. energy dependence of ${\cal P}_{\text{recon}}$
is very slow: between 150 and 200
a factor of 2. Here it is useful to remember that the W$~{\pm}$
are never produced at rest with respect to each other: the
na\"{\i}ve Breit--Wigner mass distributions are distorted
by phase-space effects, which favour lower W masses.
For 150--200
in the range 22--60 GeV, rather than in the range 0--60
It is largely this momentum that indicates how
fast the two W systems are flying apart, and therefore how much
they overlap in the middle of the event. Also the energy variation
in the perturbative description is very small. If we want to call
colour reconnection a threshold effect, we have to acknowledge that
the threshold region is very extended.

Comparing the scenarios I and II above with the no-reconnection
scenario, it turns out that reconnection effects are very small.
The change in the average charged multiplicity is at the level
of a per cent or less, and similar statements hold for
rapidity distributions, thrust distributions, and so on. This
is below the experimental precision we may expect, and
so may well go unobserved. One would like to introduce more clever
measures, which are especially sensitive to the interesting features,
but so far we have had little success.

Ultimately, the hope would be to distinguish
scenarios
into the nature of the confinement mechanism.
In principle, there are such differences. For instance,
the reconnection probability is much more sensitive to
the event topology in scenario
of having two string cores cross is more selective than
that of having two broad flux tubes overlap.

We now come to the single most critical observable for LEP
physics, namely the W mass. Experimentally, $m_{\text{W}}$ depends
in a non-trivial fashion on all particle momenta of an event.
Errors in the W mass determination come from a number of sources
\cite{LEP2work}, which we do not intend to address here. Therefore
we only study the extent to which the average reconstructed W mass
is shifted when reconnection effects are added, but everything else
is kept the same. Even so, results do depend on the reconstruction
algorithm used. We have tried a few different ones, which however
all are based on the same philosophy: a jet finder is used to define
at least four jets, events with two very nearby jets or with more
than four jets are rejected, the remaining jets are paired to
define the two W's, and the average W mass of the event is
calculated. Events where this number agrees to better than 10 GeV
with the input average mass are used to calculate the systematic
mass shift.

In scenario I this shift is consistent with being zero, within the
10
(160,000 events per scenario). Scenario
shift, of about $-30$
of the basic scheme. A simpler model, where reconnections are
always assumed to occur at the centre of the event, instead gives
a positive mass shift: about $+30$
${\cal P}_{\text{recon}} \approx 0.35$. We are therefore forced to
conclude that not even the sign of the effect can be taken for
granted, but that a real uncertainty of $\pm 30$
from our ignorance of non-perturbative reconnection effects.

To examine the perturbative rearrangement effect, we have used a
scenario where the original
$\text{q}_1 \overline{\text{q}}_2$ and
$\text{q}_3 \overline{\text{q}}_4$ dipoles
are instantaneously reconnected to
$\text{q}_1 \overline{\text{q}}_4$ and
$\text{q}_3 \overline{\text{q}}_2$ ones,
and these are allowed to radiate gluons with an upper
cut-off given by the respective dipole invariant mass \cite{GPZ}.
This gives a mass shift by about $+500$
argued that real effects would be suppressed by at least a factor
of $10~{-2}$ compared to this, and thus assign a 5
this source. Finally, the possibility of an interplay between the
perturbative and non-perturbative phases must be kept in mind.
We have no way of modelling it, but believe it will not be much
larger than the perturbative contribution, and thus assign a
further 5
independent, the numbers are added linearly to get an estimated
total uncertainty of 40

In view of the aimed-for precision, 40
non-negligible. However, remember that as a fraction of the W
mass itself it is a half a per mille error. Reconnection effects
are therefore smaller in the W mass than in many other observables,
such as the charged multiplicity.

Clearly, it is important to study how sensitive experimental
mass reconstruction algorithms are, and not just rely on the
numbers of this paper. We believe that the uncertainty can be reduced
by a suitable tuning of the algorithms, e.g. with respect to the
importance given to low-momentum particles, and with respect to the
statistical treatment of the wings of the W mass distribution.
The 40
to be assigned to any algorithm that has not been properly
evaluated.

In summary, we have developed the first detailed model of QCD
rearrangement effects in the decay of two heavy colourless objects
into quarks and gluons. Beyond the immediate use for
LEP
other processes, e.g. top quark production and decay \cite{K4,K5,KA}.


\begin{thebibliography}{99}

\bibitem[*]{VAK}
Supported by the UK Science and Engineering Research Council.

\bibitem{bigpaper}
T. Sj\"ostrand and V.A. Khoze, preprint CERN-TH.7011/93 and DTP/93/74

\bibitem{LEP2work}
LEP 2 Workshop presentation by L. Camilleri at the LEPC open meeting,
CERN, November 1992

\bibitem{GPZ}
G. Gustafson, U. Pettersson and P. Zerwas,
Phys. Lett. {\bf B209} (1988) 90

\bibitem{K4}
Yu.L. Dokshitzer, V.A. Khoze, L.H. Orr and W.J. Stirling,
Nucl. Phys. {\bf B403} (1993) 65

\bibitem{K2}
Yu.L. Dokshitzer, V.A. Khoze, A.H. Mueller and S.I. Troyan,
`Basics of Perturbative QCD', ed. J. Tran Thanh Van (Editions
Fronti\`eres, Gif-sur-Yvette, 1991)

\bibitem{K5}
V.A. Khoze, L.H. Orr and W.J. Stirling, Nucl. Phys. {\bf B378}
(1992) 413

\bibitem{K8}
Ya.I. Azimov, Yu.L. Dokshitzer, V.A. Khoze and S.I. Troyan,
Phys. Lett. {\bf 165B} (1985) 147

\bibitem{Lund}
B.
Phys. Rep. {\bf 97} (1983) 31

\bibitem{Pythia}
T. Sj\"ostrand and M. Bengtsson, Comput. Phys. Commun.
{\bf 43} (1987) 367; \\
H.-U. Bengtsson and T. Sj\"ostrand, Comput. Phys. Commun.
{\bf 46} (1987) 43; \\
T. Sj\"ostrand, preprint CERN-TH.6488/92

\bibitem{KA}
T. Sj\"ostrand and P. Zerwas, in `e$~+$e$~-$ Collisions at 500 GeV:
The Physics Potential', ed. P.M. Zerwas (DESY 92-123A, Hamburg, 1992),
p. 463

\end{thebibliography}
\end{document}